\documentclass{article}

\usepackage[utf8]{inputenc}
\usepackage{blindtext}
\usepackage[margin=1in]{geometry}

\usepackage{amsmath, amssymb, graphicx, hyperref, bm, tabu, enumitem,verbatim}
\RequirePackage[backend=biber,authordate,giveninits=true,uniquename=mininit,natbib,noibid]{biblatex-chicago}

\usepackage{booktabs,color}
\newcommand{\vect}[1]{\boldsymbol{\mathbf{#1}}}
\newcommand{\logit}{\text{logit}}

\title{Bias and Excess Variance in Election Polling: \\ A Not-So-Hidden Markov Model}
\author{Graham Tierney$^*$\blfootnote{Corresponding author. Address: 214 Old Chemistry, Box 90251, Durham, NC 27708-0251. Email: graham.tierney@duke.edu} and Alexander Volfovsky}

\newcommand\blfootnote[1]{%
\begingroup
\renewcommand\thefootnote{}\footnote{#1}%
\addtocounter{footnote}{-1}%
\endgroup
}

\providecommand{\keywords}[1]
{
  \small	
  \textbf{Keywords---} #1
}

\begin{document}

\maketitle

\begin{abstract}
   With historic misses in the 2016 and 2020 US Presidential elections, interest in measuring polling errors has increased. The most common method for measuring directional errors and non-sampling excess variability during a postmortem for an election is by assessing the difference between the poll result and election result for polls conducted within a few days of the day of the election. Analyzing such polling error data is notoriously difficult with typical models being extremely sensitive to the time between the poll and the election. We leverage hidden Markov models traditionally used for election forecasting to flexibly capture time-varying preferences and treat the election result as a peak at the typically hidden Markovian process. Our results are much less sensitive to the choice of time window, avoid conflating shifting preferences with polling error, and are more interpretable despite a highly flexible model. We demonstrate these results with data on polls from the 2004 through 2020 US Presidential elections and 1992 through 2020 US Senate elections, concluding that previously reported estimates of bias in Presidential elections were too extreme by 10\%, estimated bias in Senatorial elections was too extreme by 25\%, and excess variability estimates were also too large. 
\end{abstract}


\section{Introduction}

Election polls spur media discussion, inform candidate and voter choices, and provide inputs to election forecasts \citep{sunshine2011evolution}. Candidates use polls to allocate campaign resources and voters may rely on polls to inform strategic decisions about who to vote for and whether to turnout to vote at all \citep{huang2009beyond,fey1997stability,levine2007paradox}. Recent high-profile polling misses both in the US and the UK have called into question the accuracy of polls and of forecasts based on poll aggregations \citep{jackson2020uncertainty,kennedy2018evaluation,sturgis2016report,sturgis2018assessment}. Pollsters now have to grapple with declining response rates, changing methods of contact, and turbulent turnout dynamics, all of which make assessing who is being sampled and how to compare the sampled population to the expected voting population more difficult \citep{hillygus2016polling}. 

Knowing that errors exist, however, does not make measuring polling errors any easier. Errors can come in two forms. Polls may suffer from a directional error, consistently over- or under-estimating one candidate's support, and excess variance, variability above what would be implied if polls were independent random samples of the electorate. Estimating both of these quantities requires comparing a poll's result to the underlying value it measures. Making such comparisons is complicated by the fact that the ``ground truth'' of voters' preferences is only observed once, when the election happens, while polls measure preferences at some earlier point in time. This early measurement could be inaccurate simply due to temporal dynamics: undecided voters breaking heavily for one candidate, already-decided voters changing their opinion, or poll respondents making different turnout decisions than expected. 

Standard solutions involve only using polls conducted close to the election and assuming that preferences do not change in that time window
\citep[e.g.][]{jennings2018pollerrors}
or specifying a simple (linear) model for how preferences might change over time
\citep[e.g.][]{ShiraniMehr2018biasvariance}. While limiting the amount of polling data that enters the model can help these assumptions hold, it also risks results changing based on the amount of data used. We demonstrate that this does in fact happen: the conclusions of these methods are inconsistent across the subjective inclusion windows, i.e. the candidate whose support is overstated by polls changes based on how many days of polling are included. Moreover, when the assumptions about how preferences evolve are incorrect, these methods will mislabel changes in preferences as polling errors with high precision because they do not properly account for model misspecification and the fact that the truth and measurement are observed at different times. Finally, certain implementations require modeling polling errors on the logistic scale to ensure that relevant quantities are bounded between 0 and 1. However, this complicates interpretation of estimated polling errors because they require an inverse-logistic transformation to report results with meaningful units and the directional error varies with the actual election result. 

\begin{figure}[htbp]
    \centering
    \includegraphics[width=.90\textwidth]{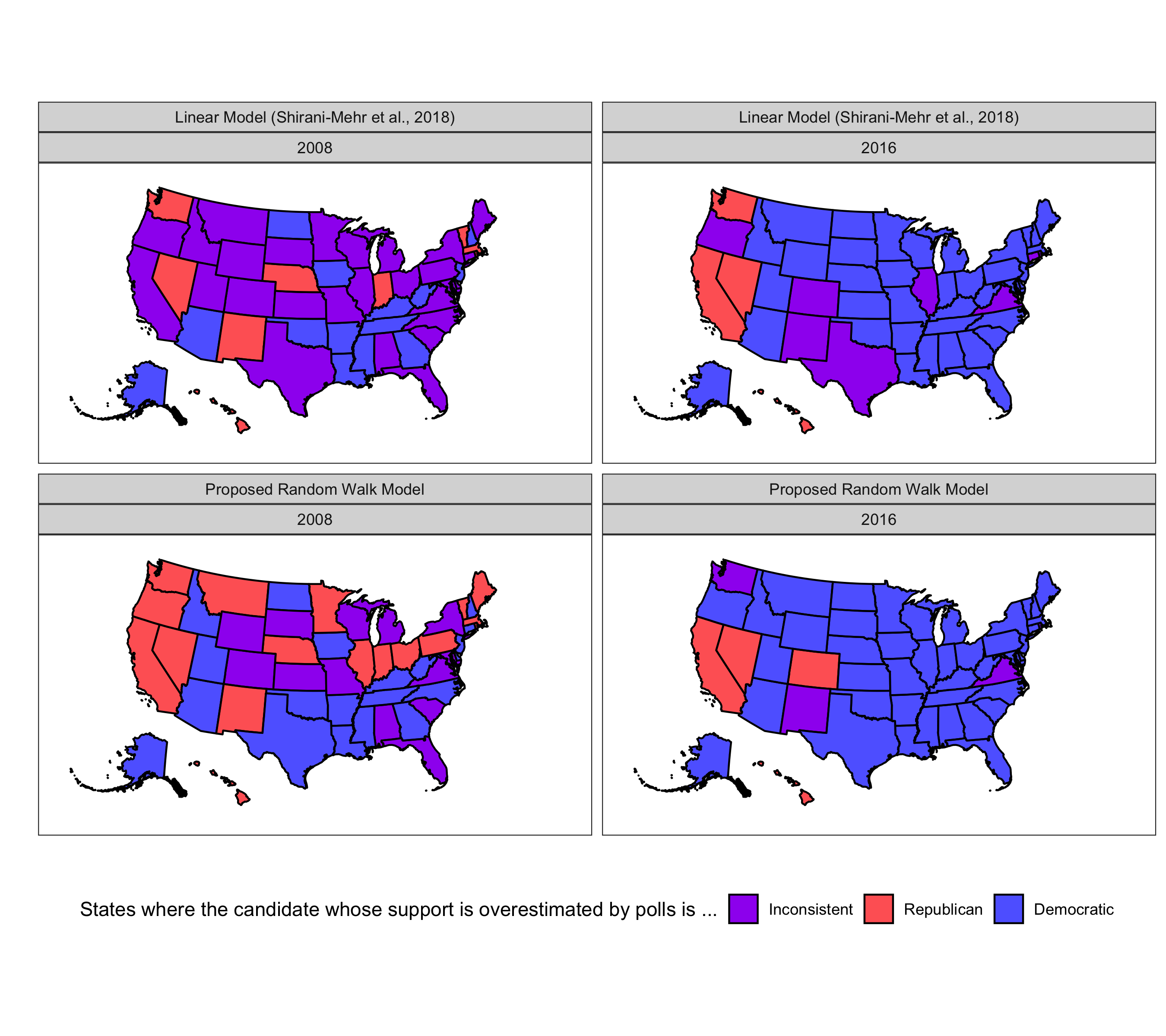}
    \caption{\textbf{Shifting Results by Inclusion Window in Presidential Elections.} Figure shows for the two most-polled election cycles in our data, 2008 and 2016, states where polling error estimated by the linear model of  \citet{ShiraniMehr2018biasvariance} and our proposed model (Section \ref{sec:model}) \textit{changed signs}. Purple colouring indicates states where models estimate that polls are biased towards different candidates depending on which polls are included. In 2008, a very turbulent year, our model estimates much more consistent errors. In 2016, when polls overstated the Democrats support, both models are usually consistent, but our model is slightly more so.}
    \label{fig:flip_elections}
\end{figure}

Our proposed solution to identifying biases in polls borrows from tools frequently used in the election forecasting literature  \citep{jackman2005pooling,Linzer2013election}. We specify a flexible hidden Markov model for how preferences change over time and treat the election outcome as a peak at the typically-hidden, underlying Markovian process. This approach directly builds upon the methodology in \citet{ShiraniMehr2018biasvariance}, and has three principle advantages: 

\begin{enumerate}
    \item Consistency across time-windows: We estimate election-level errors and excess variance \textit{without} relying on a tight time window around the election. Our model's estimates are remarkably consistent whether using data from polls conducted within 10 or 100 days of the election, while other models markedly shift their estimates of bias depending on the time window. The model learns and applies a weighting scheme to down-weight polls conducted far from the election, where the weights are determined by the variability of preferences: if preferences are highly variable then the information in a poll becomes outdated quickly, if preferences are stable then the information persists. This issue is highlighted in Figure~\ref{fig:flip_elections}, where we show which candidate's support is overstated by polls changes by inclusion window for traditional models but not our model. 
    \item Avoid conflating changes in preferences with polling error: \textit{Polls are a snapshot, not a forecast}. If preferences are shifting, then even extremely accurate polls will appear ``biased'' when compared with the election result. This and the above point are especially apparent in 2008 swing states and select 2016 Presidential races with distinct non-linear trends in preferences. Simple models for preferences will mistakenly attribute non-linear trends to directional error or excess variance, which exaggerates the failings of polling data. 
    \item Interpretability: Simple models for changes in errors and preferences, e.g. linear trends as in \citet{ShiraniMehr2018biasvariance}, require modeling bias after a transformation of key parameters to ensure polls are measuring a quantity between 0\% and 100\%. Our model for preferences is flexible enough that these kind of transformations are not necessary, which enables easier interpretation, direct modeling of bias, and results that do not change based on the actual election result. In particular, if error is modeled on the logistic scale, as it is in \citet{ShiraniMehr2018biasvariance}, then the error of a poll on election day must be computed by initially taking the logistic transformation of the election result, adding the estimated logistic error, then taking the inverse-logistic transformation, and finally subtracting the actual election result. This number changes based on the election result because of how error is modeled. That variability makes using results in forecasts or to ``debias'' current polls quite difficult. In contrast, with our method a single parameter measures polling bias directly and is invariant to changes in actual the election result. 
\end{enumerate}

We estimate that polls did not systematically over- or under-state either party across state-level contests, while they did overstate the Democrats' support by approximately 2 percentage points in 2016 and 2020 Presidential elections. In contrast to directly comparing the poll and election result or using a linear trend adjustment \citep{ShiraniMehr2018biasvariance}, we estimate less extreme errors by a factor of 10\% to 25\% and much smaller excess variances across all election cycles. Unlike previous approaches, our estimates are not sensitive to how many days of polling are included, letting us leverage a larger sample and estimate more credible bias and excess variability parameters. In particular, without a flexible model for preferences, other models conflate changes in preferences with polling error, estimating excess variability that is two to three times larger than what our model indicates.

\section{Background}\label{sec:background}

In this section we situate our methodological points within prior work on polling in Section~\ref{sec:background_polling}, documenting the potential sources of non-sampling error and standard methodologies, and prior work using hidden Markov models for election forecasting and poll aggregation in Section~\ref{sec:forecasting}, documenting common issues with such methods that our application avoids. 

\subsection{Polling Accuracy}\label{sec:background_polling}

Pre-election or ``horserace'' polls have a long history in the United States. Pollsters typically attempt to contact a representative sample from the voting population (or weight a random sample to match the expected voting population) and ask respondents which Presidential candidate they intend to vote for. While errors certainly do arise from the random sampling, the extensive literature on total survey error documents many non-sampling reasons for polling errors \citep{weisberg2009total,biemer2010total,groves2010total}. For example, non-response bias may occur when supporters of a candidate with low support may be less likely to respond to polls \citep{gelman2016mythical}. Other sources of error include order effects \citep{mcfarland1981effects} and question wording \citep{smith1987we}. For election polls specifically, many pollsters poll the same race and differences in survey methodology and question wording contribute to ``house effects'' whereby each pollster may measure preferences slightly differently. \citet{mcdermott2003horserace} study house effects in the 2000 US Presidential election and \citet{jackman2005pooling} study them in the Australian context.  Our method studies non-sampling errors to examine how they vary across election cycles and voting populations.

After the high-profile polling misses in the state-level results of the 2016 and 2020 US Presidential elections, along with misses in the 2015 UK general election and 2016 Brexit referendum referendum, practitioners began to question the relevance of election polling altogether \citep{barnes2016reality}. Indeed, lengthy retrospective reports about those elections and horse-race polls were produced, which suggest that non-representative samples and potentially late swings in opinion contributed to the errors \citep{sturgis2016report,aapor2017evaluation,sturgis2018assessment,bon2019polling,aapor2021evaluation}. 

The central theme of the literature that we build upon is that polling errors change over the course of a campaign and across different election cycles and electorates \citep{jennings2018pollerrors}. As undecided voters commit to a candidate and late-breaking news stories affect voter preferences, poll results late in the campaign are generally closer to the actual election outcome. A frequent methodological choice in this literature is to compute poll-specific errors as the difference between a poll's stated support for each candidate and the election outcome for some small time window close to the election. For example, \citet{kennedy2018evaluation} use polls within 13 days of the election and \citet{jennings2018pollerrors} use those within 7 days. This structure was relaxed most recently in \citet{ShiraniMehr2018biasvariance} who allow for linear changes throughout a 21-day period before the election. More recently, \citet{bon2019polling} used this model to decompose bias due to undecided voters from other sources with special attention to the 2016 US Presidential election, which had an unusually high number of undecided voters close to the election. We detail the \citet{ShiraniMehr2018biasvariance} method further in Section~\ref{sec:model} as a principle comparison for the model we develop. Our model will flexibly capture shifting preferences and account for them when estimating polling error. 

\subsection{Forecasting and Poll Aggregation}\label{sec:forecasting}

Over the last 20 years, aggregating polls to create more precise estimates of the electorate's preferences and forecasting eventual election results has risen in popularity \citep{jackson2018rise}. Even earlier, researches developed methods to combine polls and public opinion surveys to separate real changes in preferences from survey error and identify the impact of campaign events \citep{green1999tracking,erikson1999presidential,wlezien2002timeline}. We focus this section on only the specific class of models that we build upon for our method, highlighting key innovations and areas where our application simplifies certain assumptions. \citet{pasek2015predicting} provide a detailed review of alternative methods for election forecasting and poll aggregation beyond the hidden Markov approach we focus on here. 

\citet{Linzer2013election} developed a hidden Markov model for US Presidential election forecasting where underlying state-level preferences evolve over time following a random walk. A Bayesian estimation procedure enables forecasting Electoral College outcomes via posterior predictive simulations. \citet{jackman2005pooling} outlines a very similar model that is more focused on pooling polls to estimate current preferences and house effects rather than making explicit election forecasts. \citet{pickup2007campaign,pickup2008campaign} expanded upon Jackman's work to estimate house effects and industry-wide bias in the 2004 and 2006 Canadian and 2004 US Presidential elections. Our method, rather than estimating pollster-specific errors from multiple polls of the same race, will estimate aggregate errors across multiple elections. 

The underlying principle, that polls are noisy measurements of latent preferences that change over time, was developed for general public opinion tracking in \citet{green1999tracking}, and has been expanded and applied by many forecasting models, including multi-party systems \citep{walther2015picking,stoetzer2019forecasting}, and the Economist's 2020 forecast, which made additional adjustments for correlated shifts in state-level trends and polling errors \citep{heidemanns2020updated}. These models often ``debias'' current polls by correcting for historical polling errors \citep{rothschild2009forecasting}. Recent work applied this general principle to predicting US Senatorial elections with more complex mapping from polling data to election outcomes \citep{chen2022polls}. The general modeling framework whereby surveys measure latent preferences has been applied beyond election polls \citep[e.g.][]{caughey_warshaw_2015}. 

These methods typically incorporate ``fundamentals,'' historical data on election outcomes, broad economic and political features, and potentially very early polls, into priors on election results \citep{abramowitz2008forecasting,campbell1990trial,erikson2008leading}. We, however, will condition on the election outcome to estimate polling errors, removing the need for forecasting and applying fundamentals approaches altogether. Another area of concern for these models is how correlated changes in preferences are across electoral units. Consistent with \citet{ShiraniMehr2018biasvariance}, we allow sharing of information on the election-level parameters through a hierarchical model across elections. 

Section~\ref{sec:background_polling}'s discussion of non-sampling sources of error highlight that polls have significant sources of unknown uncertainty, which contribute to forecasts based on those polls making overconfident and inaccurate predictions \citep{jackson2020uncertainty}. The overconfidence is largely attributable to the fact that the (invalid) assumptions that polls are independent from each other and random samples from the electorate lead to significantly underestimating the variance of estimators that combine results from multiple polls \citep{clinton2013robo}. While the model we use stems from the forecasting literature, we will treat the election outcome as known and use that information to derive estimates of historic polling errors and uncertainty, which can inform and improve future forecasts. 

\section{Data}

Our data consist of the 100 day sample of polls used in \citet{ShiraniMehr2018biasvariance} covering 2004 through 2012 elections and is supplemented with 2016 and 2020 polling data for the 100 days preceding those elections collected by the Economist for use in their forecasting model. We also rely on the Senate polls collected in \cite{chen2022polls}, who use them to train an election forecasting model. The Senate polls cover elections from 1992 to 2020. All data and code are available at this \href{https://github.com/g-tierney/polling_errors_replication}{GitHub repository}. 

\begin{figure}[ht]
    \centering
    \includegraphics[width=.7\textwidth]{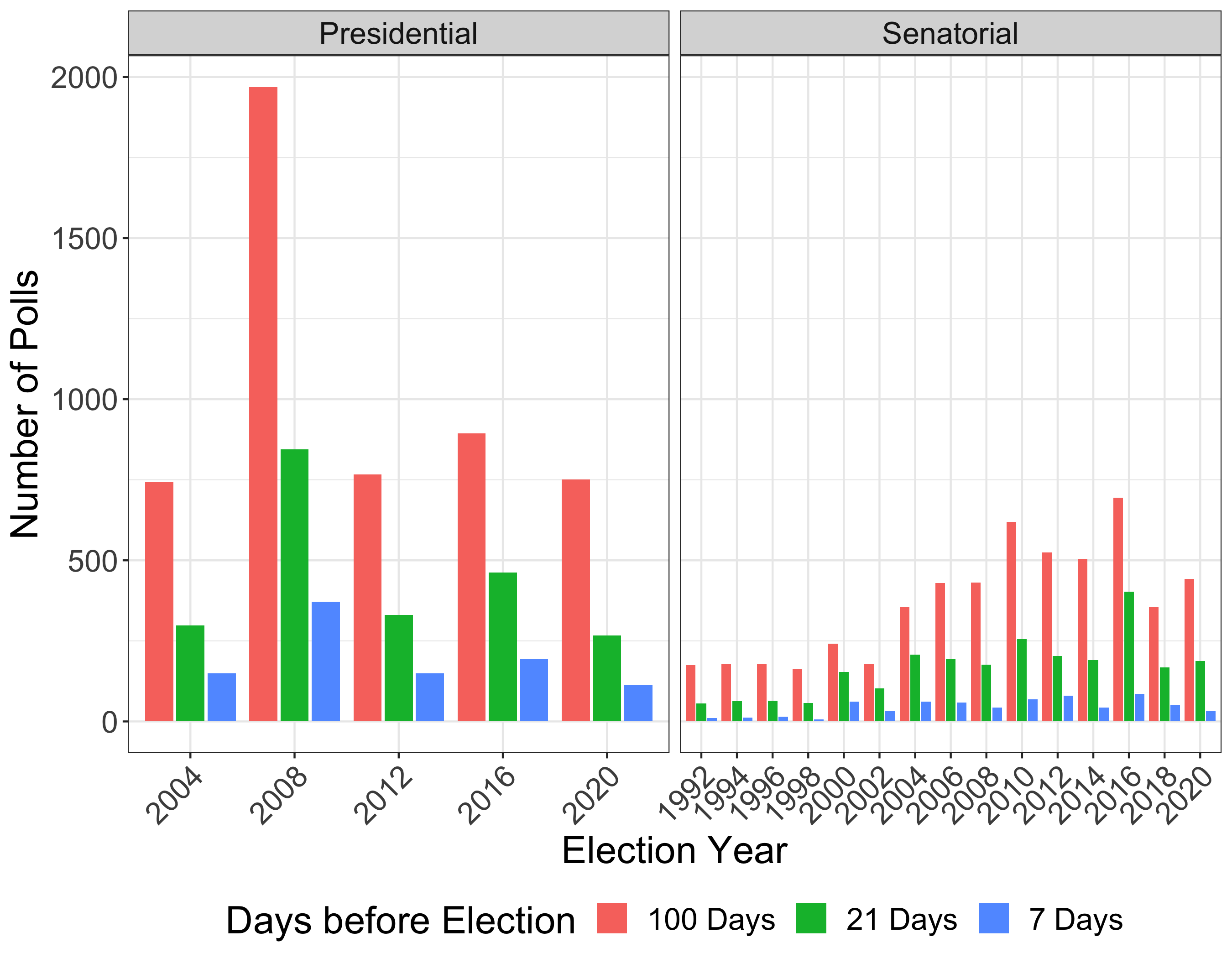}
    \caption{\textbf{Number of Polls by Varying Time Window.} Columns show the number of polls conducted each election cycle; colours indicate the cutoff time. Polls are conducted more frequently towards the end of the election cycle; approximately 19\% and 43\% of polls conducted in the final 100 days are conducted in the final 7 and 21 days, respectively.}
    \label{fig:npolls_by_year}
\end{figure}

\begin{figure}[ht]
    \centering
    \includegraphics[height=.45\textheight]{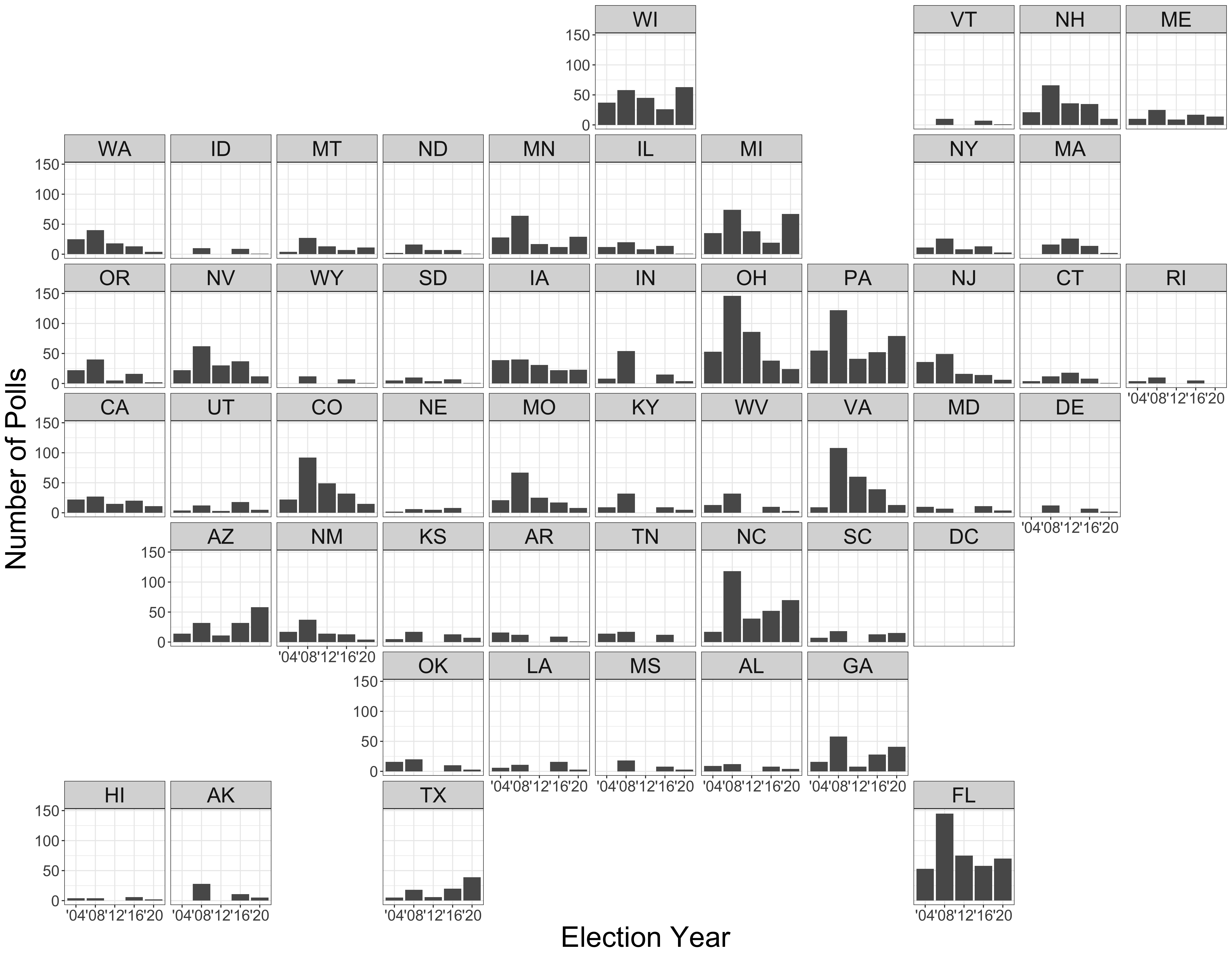}
    \caption{\textbf{Number of Presidential Polls by State and Year.} Columns show the total number of polls conducted within 100 days of the election for each election cycle in each state. Swing states are polled much more frequently than non-swing states, and many states are not polled at all in some election cycles.}
    \label{fig:npolls_by_year_state}
\end{figure}

The availability of polling data varies across elections and states. Elections years and swing states where the outcome is genuinely in doubt are polled more frequently. Figure~\ref{fig:npolls_by_year} shows the number of polls by election year for varying time windows before the election. The election in 2008 was the first time when an African American candidate was nominated by a major political party, and more polls were conducted that year than in any other. Senatorial elections are polled less frequently than Presidential elections, but there has been a notable increase in the number of polls over time. Figure~\ref{fig:npolls_by_year_state} shows the number of polls conducted at most 100 days before the election by state and year for Presidential contests. The 2008 election cycle again stands out, as do many swing states such as Ohio, Pennsylvania, and Florida. We do not show a similar map for Senatorial elections because of the timing irregularity of Senate contests; not every state has a Senate race in every cycle. 

\section{Proposed and Comparison Models}\label{sec:model}

The core model that we study is that poll $i$ of election $r_i$ conducted $t_i$ days before the election reports $y_i$, the proportion of the sample that intend to vote for the Republican candidate out of people who intend to vote for either the Republican or the Democrat (omitting third-party and undecided voters), and $n_i$, the number of intended two-party voters. In the election, the Republican's portion of the two-party vote is $v_{r_i}$. Some polls are conducted years before the actual election, so all models require a cutoff day $T$ that identifies which polls to look at, i.e. only polls with $t_i \leq T$. Below we discuss several comparison models and develop our proposed model. While all models share the assumption $y_i \sim N(p_i,\sigma^2_i)$, \textit{they differ in how $p_i$, true underlying preferences measured by poll $i$, is decomposed}.

\textbf{M1: Static Model}. The simplest model we consider is commonly used in practice where $p_i := v_{r_i} + \alpha_{r_i}$. Writing $y_i - v_{r_i} \sim N(\alpha_{r_i},\sigma^2_i)$, this assumes that the electorate's preferences do not change over time and $\alpha_{r_i}$ is a time-invariant, election-specific error. This requires choosing a small time cutoff $T$ ``close'' to the election to justify the assumption of static preferences. 

\textbf{M2: Linear Model} The model in \cite{ShiraniMehr2018biasvariance} sets  $\logit(p_i) = \logit(v_{r_i}) + \alpha_{r_i} + \beta_{r_i}t_i$ and is our principle comparison model. The authors refer to the sum $\alpha_{r_i} + \beta_{r_i}t_i$ as ``error.'' An equivalent interpretation that clarifies the comparison with our model is to interpret $\logit(v_{r_i}) + \beta_{r_i}t_i$ as preferences that change linearly over time, and thus $\alpha_{r_i}$ is the only ``error'' term. ``Error'' is used here rather than bias because, due to the logistic transformation, the $\alpha$ terms do not measure bias in the statistical sense. Note that $t_i=0$ corresponds to a poll conducted on the day of the election, so regardless of whether one thinks of $\beta_{r_i}t_i$ as preferences or error, the ``election day error'' is measured with $\alpha_{r_i}$ alone. Also note that the logit transform is necessary to ensure that $p_i$ lies between 0 and 1. A linear trend could easily imply that the expected poll proportion is outside of $[0,1]$. 

\textbf{M3: Random Walk (RW) Model.} This is our proposed model where we set $p_i = \theta_{r_i,t_i} + \alpha_{r_i}$. $\theta_{rt}$ represents the electorate's preferences at time $t$ and evolves via a reverse random walk: $\theta_{r,t+1} \sim N(\theta_{rt},\gamma_r^2)$. The election result is the reveal of $\theta_{r,0} := v_{r}$. Note the lack of a logit transform on $p_i$. While one could include it, the flexibility of the random walk and small estimated $\gamma_r$ terms mean that in practice neither the parameter $\theta_{rt}$ nor $\theta_{rt} + \alpha_t$ come close to leaving the interval $[0,1]$. When such models are used for forecasting, the logistic transformation is more common, as in \cite{Linzer2013election}, because $v_r$ is unknown and forecasts are made with long time horizons. Our model conditions on $v_r$ so does not need the transformation. A detailed discussion of this model is presented in the next section.  

\textbf{Variance terms.} The difference in the above models is in the specification of each poll's expected value $p_i$, but the variance term $\sigma^2_i$ deserves discussion as well. The key modeling decision is whether to include the binomial variance term and, if so, how to allow for excess variance. If a poll is truly a random sample, then $\text{Var}(y_i|p_i) = p_i(1-p_i)/n_i$. It is well known, however, that election polls have higher variance than what this would imply, despite polling firms typically constructing error estimates with this assumption. Consistent with \cite{ShiraniMehr2018biasvariance}, we model this term as $\sigma^2_i = \frac{p_i(1-p_i)}{n_i} + \tau_{r_i}$. Additive excess variance is preferable to multiplicative variance because a poll's error cannot be shrunk to essentially zero with large enough sample size. National and online-only polls can have quite large $n_i$, which makes the binomial variance shrink to near zero even for $p_i=0.50$. 

\subsection{Random Walk Model Construction}

The Static Model is quite simple and the Linear Model is detailed extensively in \citet{ShiraniMehr2018biasvariance}. Here we provide additional details and expand on the interpretaiblity of our RW Model (M3). For clarity, we formally state the model. 
\begin{align}
    y_i &\sim N\left(p_i,\frac{p_i(1-p_i)}{n_i} + \tau^2_{r_i}\right) \label{eq:y}\\
    p_i &= \min(\max(0,(\theta_{r_i,t_i} + \alpha_{r_i})),1) \label{eq:p}\\
    \theta_{r,t+1} &\sim N(\theta_{r,t},\gamma^2_r) \\
    \theta_{r,0} &:= v_{r},
\end{align}
where $\tau_r$ is the election-specific variance above simple random sampling. Similarly, $\gamma_r$ measures how much the electorate's preferences change day to day. Under this model specification, approximately 95\% of daily shifts will be $\pm 2 \gamma_r$ percentage points. Lastly, $\alpha_r$ directly measures poll bias. A poll conducted on election day ($t_i = 0$) would in expectation be off by $\alpha_r$ percentage points. The minimum and maximum operators in (\ref{eq:p}) ensure that $p_i$ lies in the interval 0 to 1 so that the variance in (\ref{eq:y}) is always non-negative. In practice, this restriction is only necessary so that early (pre-convergence) MCMC draws do not break the sampler. Posterior samples are never observed close to this boundary condition. In the Linear Model, a poll conducted on election day will have expected value $\logit^{-1}(\logit(v_r) + \alpha_r)$, and as such will have ``election day error'' of $100(\logit^{-1}(\logit(v_r) + \alpha_r) - v_r)$ percentage points. Note that this error changes depending on the actual election result.

Election-specific scalar parameters $\tau_r$, $\alpha_r$, and $\gamma_r$ have hierarchical normal or half-normal (for variance terms) priors placed on them to borrow strength across elections. $\alpha_r \sim N(\mu_\alpha,\sigma^2_\alpha)$, $\tau^2_r \sim N_+(0,\sigma^2_\tau)$, and $\gamma_r \sim N_+(0,\sigma_\gamma)$. We use the same ``weakly informative'' priors on the hyperparameters of those distributions as in \cite{ShiraniMehr2018biasvariance} for $\alpha_r$ and $\tau^2_r$ with slight modifications to account for the lack of a logistic transformation: $\mu_\alpha \sim N(0,0.05^2)$, $\sigma_\alpha \sim N_+(0,0.2^2)$, $\sigma_\tau \sim N(0,0.05^2)$. 
For $\gamma_r$, we use $\sigma_\gamma \sim N_+(0,0.01^2)$, which places nearly all prior mass on preferences changing by at most $\pm2$ percentage points. The model is estimated in Stan using Hamiltonian Monte Carlo \citep{rstan}. We also improve convergence by reparameterizing the model to let $z_{rt} := \theta_{rt} + \alpha_t$, sample the posterior of $z_{rt}$ and $\alpha_r$, then use those samples to recover $\theta_{rt}$. This is analogous to the centered parameterization described in \citet{prado2010time}. 

\subsection{Use of observed errors $y_i - v_r$}

In this section, we discus how each model estimates $\alpha_r$ as a function of observed polling errors $y_i-v_r$. We will show that our model can be thought of as a temporally weighted average of observed polling errors, whereas the Static Model is an equally weighted average that ignores temporal information and the Linear Model does not have any such representation. Consider a single election $r$ and all of its corresponding polls, $\vect Y$, conducted at least $T$ days before the election. 
For clarity of exposition, assume that $n_i$ is sufficiently large such that the binomial component of the variance $(p_i(1-p_i)/n_i)$ is negligible.

\textbf{Static Model.} A nice feature of the Static Model is that one can easily see how $y_i - v_{r}$ is used in estimation: under the Static Model, the posterior of $\alpha_r$ with a normal prior $\alpha_r \sim N(\mu_\alpha,\sigma^2_\alpha)$ is:  
\begin{equation}
    \alpha_r|\vect Y,M_1 \sim N\left(\frac{n_r/\tau_r^2}{n_r/\tau_r^2 + \sigma_\alpha^{-2}}\frac{\sum_i(y_i - v_r)}{n_r} + \frac{\sigma_\alpha^{-2}}{n_r/\tau_r^2 + \sigma_\alpha^{-2}} \mu_\alpha,\left(n_r/\tau_r^2 + \sigma_\alpha^{-2}\right)^{-1}\right),
\end{equation}
where $\vect Y$ is all the polls of election $r$ and $n_r$ is the number of election $r$ polls. With $\mu_\alpha = 0$ and $\sigma_\alpha^2$ sufficiently large, this posterior simplifies to have expected value of $\sum_i (y_i - v_r)/n_r$, an equally weighted average of the observed differences between the polls and the election outcome. Clear in this construction is the importance of $T$ under the Static Model. All polls are weighted equally, regardless of when they were conducted, so $T$ must be carefully chosen.

\textbf{Linear Model.} The Linear Model does not have this interpretation even when $\beta_r = 0$. Recall that under this model for a poll with large sample size: $y_i \sim N(p_i,\tau_r^2)$ where $p_i = \logit^{-1}(\alpha_r + \beta_r t_i + \logit(v_r))$. This likelihood is not log-quadratic in $\alpha_r$, so it is not conjugate with a normal prior. The logistic transformation required to ensure polls' expected values lie between 0 and 1 means that the estimated $\alpha_r$ cannot be expressed as a function of the observed errors $y_i - v_r$. Moreover, because of the linear assumption, additional polls showing large support for candidate A when the election result was close to 50-50, can actually cause the Linear Model to estimate that polls favor candidate B. The linear trend needs to become steeper to fit the early polls, which shifts the intercept $\alpha_r$ in the opposite direction of what the new data indicate. We demonstrate that this phenomenon does indeed happen in Section~\ref{sec:comparisons}.

\textbf{Random Walk Model.} The proposed Random Walk Model does have an intuitive use of observed polling errors. First, consider a single poll, $y_i$. Note that $\theta_{r{t_i}}$ has marginal distribution $N(v_r,t_i \gamma_r^2)$ when integrating out $\theta_{r1}$ through $\theta_{r,t_i-1}$, so $y_i \sim N(v_r + \alpha_r,t_i\gamma_r^2 + \tau_r^2)$. Thus, the posterior for $\alpha_r$ with the same normal prior as above will be:
\begin{equation}
    \alpha_r|y_i,\tau_r,\gamma_r,M_3 \sim N \left( w_i(y_i-v_r) + (1-w_i) \mu_\alpha,\lambda_i^{-1} \right),
\end{equation}
with $\lambda_i = (\tau_r^2+t_i\gamma_r^2)^{-1} + \sigma_\alpha^{-2}$ and $w_i = (\tau_r^2+t_i\gamma_r^2)^{-1}/\lambda_i$. Thus, $w_i$ is the weight given to the observed error and $1-w_i$ the weight given to the prior mean. A poll farther from the day of the election (larger $t_i$) will have less weight than one close to the election. As the electorate's preferences become more variable (larger $\gamma_r$) polls get down-weighted more the farther out they are conducted. 

These weights highlight important improvements in our model over the Static and Linear Models. If preferences are fairly constant ($\gamma_r$ is small), then polls early in the campaign still provide accurate information about $\alpha_r$ and correspondingly $w_i$ is still large even for large $t_i$. If preferences are highly variable ($\gamma_r$ is large), then early polls do not provide much information and $w_i$ will be small for large $t_i$. The Static Model makes no account for the time a poll was conducted, and the Linear Model only accounts for time with a linear trend, which does not have this dynamic weighting structure. When either alternative model's key assumption about how preferences evolve hold, our model is flexible enough to recover that same structure. When those assumptions are incorrect, our model adapts to the data and still produces valid estimates. 

We can derive analogous results when multiple polls are conducted. Integrating out $\theta_{rt}$ will induce dependence between the polls. The data contribution to the posterior mean is still a weighted average of $y_i - v_r$ across $i$, but the weights are more complex than just observation error and time until the election because of the dependency. Recall that $\vect Y$ contains all polls $y_i$ of election $r$ and $t_i$ denote the number of days before the election that poll $i$ was conducted: 
$\vect Y \sim N(v \vect 1 + \alpha \vect 1,\boldsymbol{\Sigma})$ where $\boldsymbol{\Sigma} =  \tau^2 \vect I + \gamma^2\vect T $ and $T_{ij} = \min(t_i,t_j)$. Under $p(\alpha) \propto 1$ we have 
\begin{align}
    p(\alpha|\vect Y) &\propto \exp\left(-\frac{1}{2}(\vect Y - v \vect 1 - \alpha \vect 1)'\boldsymbol{\Sigma}^{-1}(\vect Y - v \vect 1 - \alpha \vect 1)\right) \implies \\
    \alpha|\vect Y &\sim N\left((\vect Y - v\vect 1)'\boldsymbol{\Sigma}^{-1}\vect 1/(\vect 1'\boldsymbol{\Sigma}^{-1} \vect 1),\left(\vect 1' \boldsymbol{\Sigma}^{-1} \vect 1\right)^{-1}\right)
\end{align}

That is, the posterior mean of $\alpha$ is a weighted average of the observed polling errors, $y_i - v_r$ where the weights are determined by the time the poll was conducted and the variability of preferences. Note that if $\gamma_r = 0$, then $\boldsymbol{\Sigma} = \tau_r^2 \vect I$ and the result matches the Static Model. 

\section{Comparisons}\label{sec:comparisons}

We compare the models based on election day error ($\alpha_r$ for the Static and Random Walk (RW) Models and $\logit^{-1}(\logit(v_r) + \alpha_r) - v_r$ for the Linear Model) and excess margin of error ($2\tau_r$, the margin of error for poll large enough that the binomial variance is negligible). Positive values of election day error indicate the Republican candidate's support is overstated by polls. 
We estimate each model 60 times increasing the time cutoff $T$ by one day increments between 1 and 50 days before the election and five day increments between 50 and 100 days before the election, i.e. $T \in \{1,2,\ldots,49,50,55,60,\ldots,95,100\}$. 

\begin{figure}[ht]
    \centering
    \includegraphics[width=.7\textwidth]{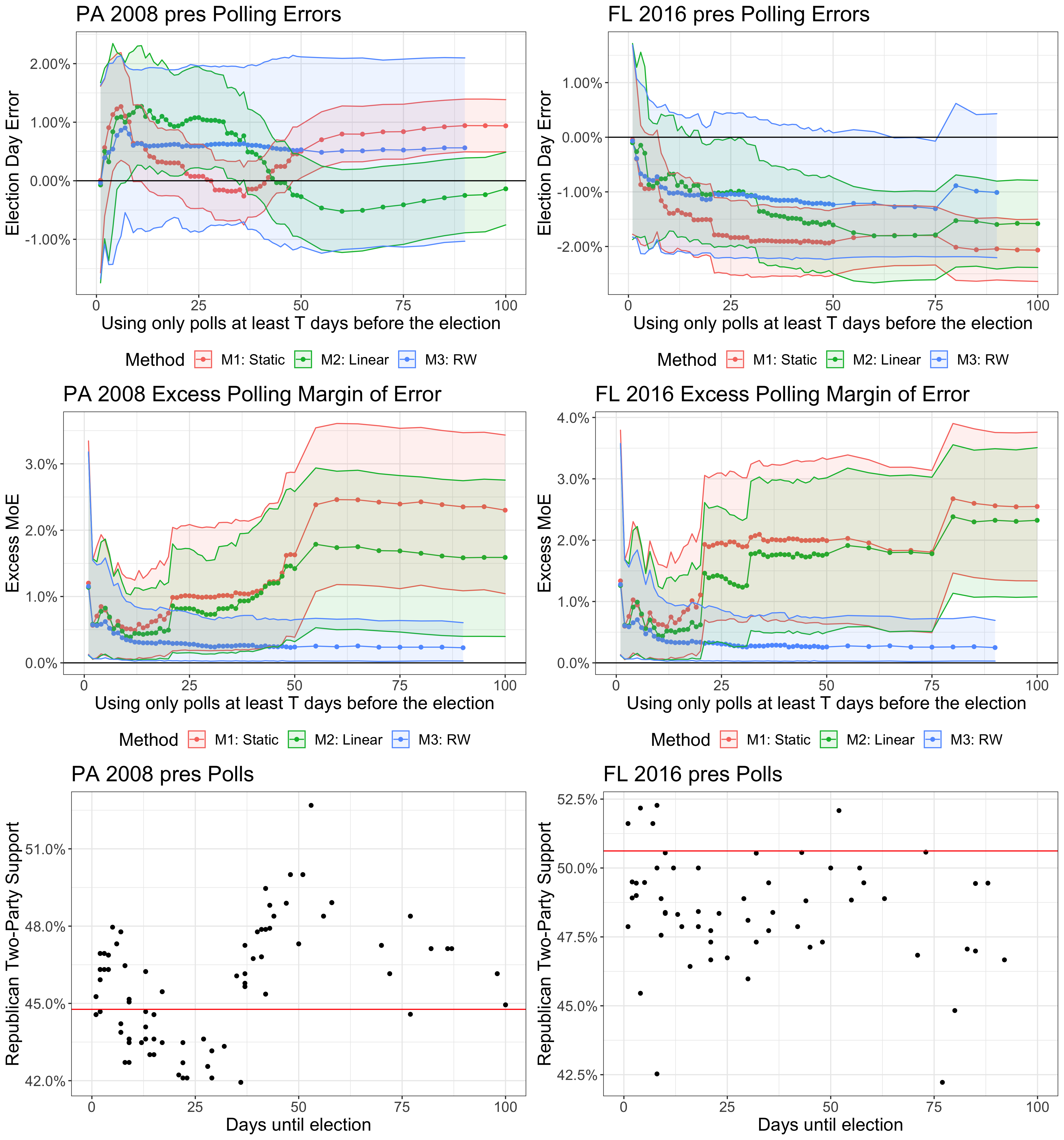}
    \caption{\textbf{Estimated Polling Error and Excess Variability by Inclusion Window for Select Elections.} Election Day Error 
    measures the expected overestimate of Republican support for a poll conducted on election day. Shaded regions show 95\% credible intervals. The bottom row shows the underlying polling data; the red line indicates the election result. Both the Static and Linear Models are sensitive to which polls are included. When their assumptions about preferences are incorrect, evidenced by the non-linearity in polls, those models increase their estimates of excess sampling variably to compensate. Our Random Walk Model, with its flexible model for preferences, does not make such erroneous estimates. }
    \label{fig:example_biases}
\end{figure}

Figure \ref{fig:example_biases} highlights two Presidential elections where the three models give different results. In the 2008 election in Pennsylvania (left panels of Figure~\ref{fig:example_biases}), many early polls showed large Republican support and large swings are evident in the 100-day period before the election. Under our proposed RW Model, election day error and excess margin of error point estimates are consistent for varying $T$. In contrast to this stability, as $T$ increases, the Static Model estimates significant positive errors and the Linear Model estimates significant negative errors. Both are trying to account for the additional polls showing large support for the Republican early in the campaign. Both alternative Models estimate much larger excess variability to compensate for the apparent misspecification of the electorate's preferences. The additional polls showing large support for the Republican candidate require that the linear trend in Model 2 slope steeply downward, which means that the \citet{ShiraniMehr2018biasvariance} model estimates that the polls likely overstate the \textit{Democrat's} support when polls showing broad support for the \textit{Republican} are added to the sample. Our Random Walk Model avoids this issue with its flexible, non-linear model for preferences and higher weight given to polls close to the election. Florida in 2016 shows a similar but less extreme example (right panels of Figure~\ref{fig:example_biases}). There was a late shift in support towards candidate Trump, and polls very close to the election were fairly accurate. All models estimate 95\% CIs that include zero for $T=10$, but as more polls are added, both Static and Linear Models become increasingly confident that the polls overstated candidate Clinton's support, while our model weights the accurate polls conducted close to the election higher and avoids making that same mistake.

\begin{figure}[!htbp]
    \centering
    \includegraphics[width=.95\textwidth]{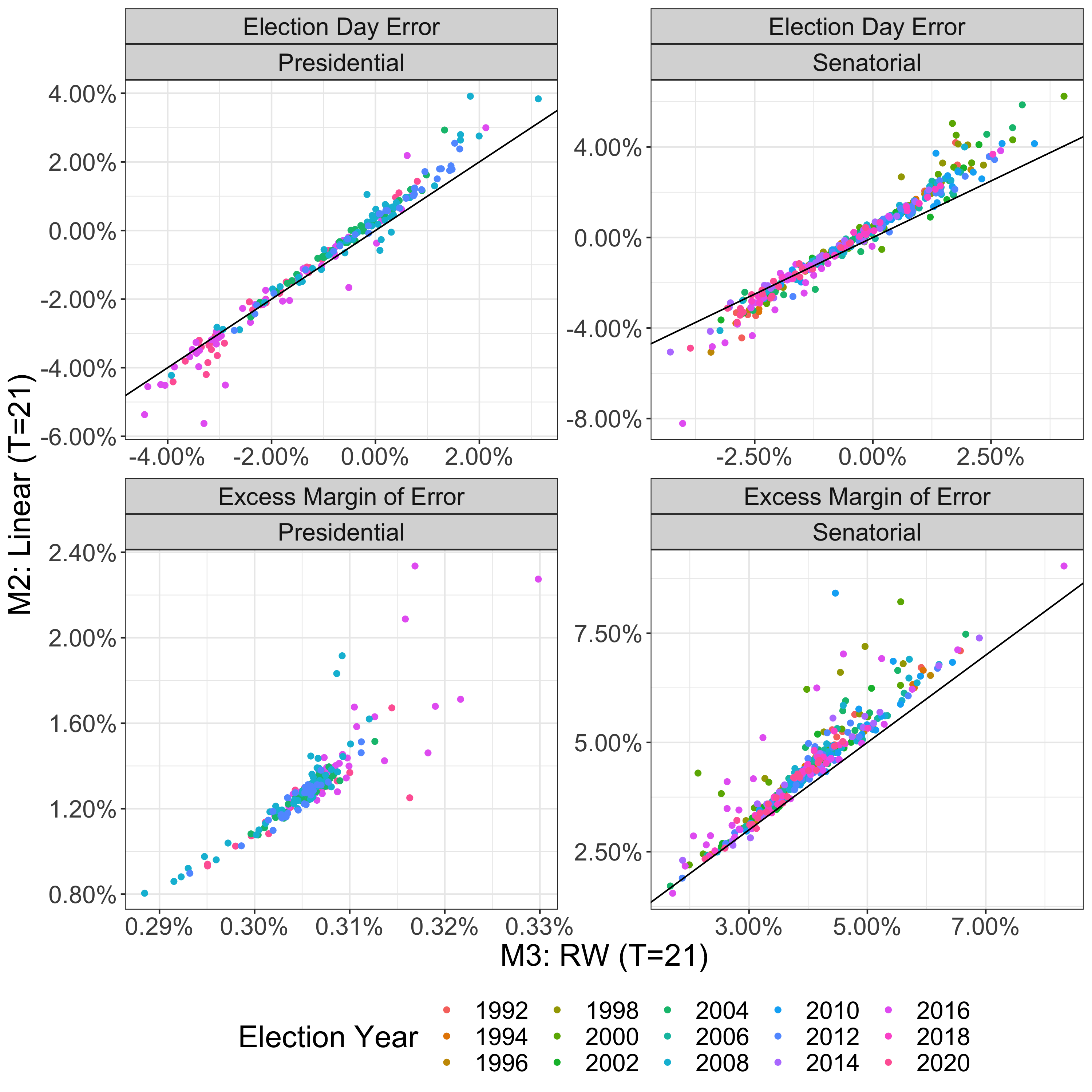}
    \caption{\textbf{Comparison of Election Day Error and Excess MoE Estimates for the Linear (M2) and Random Walk (M3) Models.} Each point compares either the election day error (top row) or excess margin of error (bottom row) for Presidential (left column) or Senatorial (right column) elections. All polls conducted within 21 days of the election are used, matching the implementation in \citet{ShiraniMehr2018biasvariance}. The Linear Model estimates more extreme biases and larger margins of error than the more flexible Random Walk Model. }
    \label{fig:m2_m3_comp}
\end{figure}

Next, we compare the models' election day error and excess variance estimates across all contests in our data using $T=21$ to match the implementation in \cite{ShiraniMehr2018biasvariance}. Because the Static Model is essentially a special case of both the Linear Model ($\beta_r = 0$) and of the RW Model ($\gamma_r = 0$), we focus our comparison on the Linear and RW Models. Figure \ref{fig:m2_m3_comp}  plots the election day error (top row) and excess margin of error (bottom row) for Presidential (left column) and Senatorial (right column) elections with colours indicating the election year. 

The Linear Model consistently estimates more extreme election day error than the Random Walk Model across both kinds of elections. Both models indicate that polls overestimated Secretary Clinton and President Biden's support in 2016 and 2020 (negative errors) but the Linear Model's estimates are larger in magnitude. A similar pattern holds for contests with positive errors when polls overstate the Republican's support: the Linear Model estimates larger errors than our Random Walk model. This is directly attributable to the model assumptions. \textit{Any} non-linear change in preferences in the three weeks before the election must be directional polling error in the Linear Model by assumption, whereas our more flexible approach avoids that mistake. 

The Linear Model also estimates much larger $\tau_r$ terms, especially in Presidential elections where the unity line cannot even be shown on the plot. The Linear Model indicates that a Presidential poll large enough to ignore binomial sampling variability should still report a margin of error about 2 to 3 times larger than the margin of error estimated by the RW Model. In Senatorial elections, both models estimate much larger excess margin of errors than in Presidential elections. The Linear estimates are almost universally larger but the difference is not as extreme. Without much flexibility for measuring how preferences evolve, the Linear Model attributes violations of the linearity assumption to excess polling variability. By  allowing for any kind of temporal evolution in preferences, our RW Model estimates notably smaller variance terms. 

It is important to note that the election day error result (that the Linear Model estimates more extreme errors) is not an artifact of the \citet{ShiraniMehr2018biasvariance} $T=21$ day cutoff time, state-level dynamics, or cycle-specific features. 
Table~\ref{tab:pres_errors} shows results from regressing the Linear Model's expected election day error on our Random Walk Model's estimate with varying fixed effects for state, election year, and cutoff time. We include estimates for all elections and all cutoff times between 14 and 28 days before the election. The Linear errors are consistently about 10\% more extreme and about half of a percentage point larger than our model's estimates. Regression coefficients are all significantly different from 1, which would imply equality, with $p<0.001$. Table~\ref{tab:sen_errors} reports the same analysis for Senatorial elections. While the constant is approximately the same, Linear errors are about half a percentage point larger, the Random Walk coefficient is much larger. Linear errors are about 25\% more extreme even after controlling for sate, year, and cutoff fixed effects. Again, all coefficients in every specification are significantly different from 1 with $p<0.001$.

\begin{table}[!htbp] \centering 
  
\begin{tabular}{@{\extracolsep{5pt}}lcccc} 
\\[-1.8ex]\hline 
\hline \\[-1.8ex] 
\\[-1.8ex] & (1) & (2) & (3) & (4)\\ 
\hline \\[-1.8ex] 
 Random Walk Error & 1.108$^{***}$ & 1.108$^{***}$ & 1.122$^{***}$ & 1.129$^{***}$ \\ 
  & (0.004) & (0.004) & (0.003) & (0.003) \\ 
  & & & & \\ 
 Constant & 0.568$^{***}$ & 0.459$^{***}$ & 0.399$^{***}$ & 0.283$^{***}$ \\ 
  & (0.041) & (0.038) & (0.037) & (0.006) \\ 
  & & & & \\ 
\hline \\[-1.8ex] 
State & Yes & Yes & Yes & No \\ 
Election Year & Yes & Yes & No & No \\ 
Cutoff Time & Yes & No & No & No \\ 
\hline \\[-1.8ex] 
Observations & 3,186 & 3,186 & 3,186 & 3,186 \\ 
R$^{2}$ & 0.983 & 0.983 & 0.982 & 0.977 \\ 
Adjusted R$^{2}$ & 0.983 & 0.982 & 0.982 & 0.977 \\ 
Residual Std. Error & 0.237 & 0.242 & 0.245 & 0.277 \\ 
F Statistic & 2,697.331$^{***}$ & 3,275.191$^{***}$ & 3,424.373$^{***}$ & 134,007.400$^{***}$ \\ 
\hline 
\hline \\[-1.8ex] 
\textit{Note:}  & \multicolumn{4}{r}{$^{*}$p$<$0.05; $^{**}$p$<$0.01; $^{***}$p$<$0.001} \\ 
\end{tabular}
\caption{\textbf{Regression Analysis of Presidential Polling Errors.} Each column reports results from regressing the Linear Model's estimated error on the Random Walk Model's error for the same state, year, and cutoff time. The first column includes fixed effects for all three identifiers and subsequent columns remove them. Data are all estimated errors for Presidential elections with cutoff times between 14 and 28 days before the election.} 
  \label{tab:pres_errors} 
\end{table} 

\begin{table}[!htbp] \centering 

\begin{tabular}{@{\extracolsep{5pt}}lcccc} 
\\[-1.8ex]\hline 
\hline \\[-1.8ex] 
\\[-1.8ex] & (1) & (2) & (3) & (4)\\ 
\hline \\[-1.8ex] 
 Random Walk Error & 1.248$^{***}$ & 1.248$^{***}$ & 1.256$^{***}$ & 1.260$^{***}$ \\ 
  & (0.005) & (0.005) & (0.005) & (0.005) \\ 
  & & & & \\ 
 Constant & 0.508$^{***}$ & 0.405$^{***}$ & 0.497$^{***}$ & 0.433$^{***}$ \\ 
  & (0.081) & (0.080) & (0.079) & (0.007) \\ 
  & & & & \\ 
\hline \\[-1.8ex] 
State & Yes & Yes & Yes & No \\ 
Election Year & Yes & Yes & No & No \\ 
Cutoff Time & Yes & No & No & No \\ 
\hline \\[-1.8ex] 
Observations & 2,592 & 2,592 & 2,592 & 2,592 \\ 
R$^{2}$ & 0.977 & 0.975 & 0.973 & 0.965 \\ 
Adjusted R$^{2}$ & 0.976 & 0.975 & 0.973 & 0.965 \\ 
Residual Std. Error & 0.288 & 0.296 & 0.306 & 0.348 \\ 
F Statistic & 1,500.116$^{***}$ & 1,779.242$^{***}$ & 1,861.689$^{***}$ & 71,411.690$^{***}$ \\ 
\hline 
\hline \\[-1.8ex] 
\textit{Note:}  & \multicolumn{4}{r}{$^{*}$p$<$0.05; $^{**}$p$<$0.01; $^{***}$p$<$0.001} \\ 
\end{tabular} 
\caption{\textbf{Regression Analysis of Senatorial Polling Errors.} Each column reports results from regressing the Linear Model's estimated error on the Random Walk Model's error for the same state, year, and cutoff time. The first column includes fixed effects for all three identifiers and subsequent columns remove them. Data are all estimated errors for Senatorial elections with cutoff times between 14 and 28 days before the election.} 
  \label{tab:sen_errors} 
\end{table} 

\begin{figure}[ht]
    \centering
    \includegraphics[width=.75\textwidth]{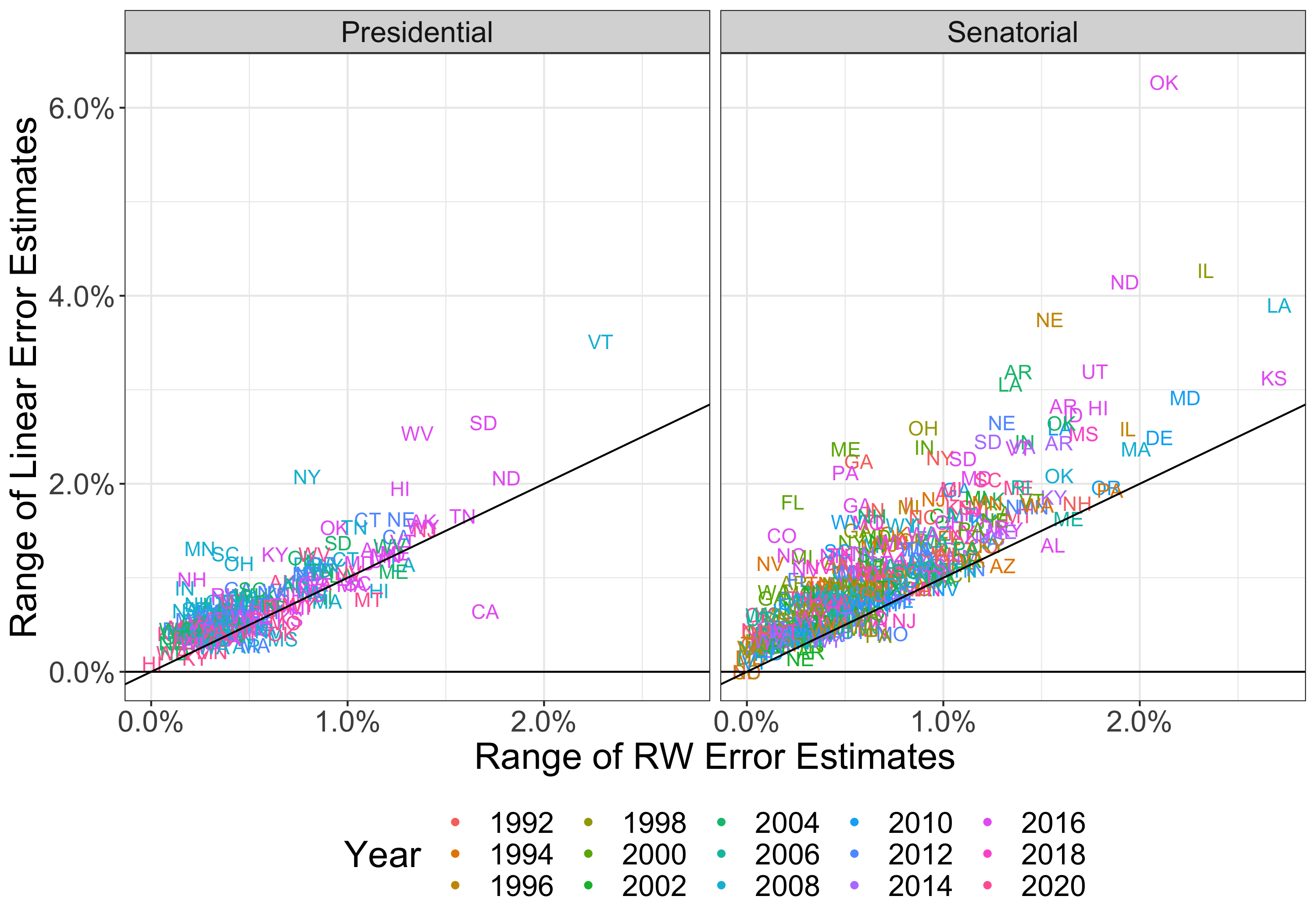}
    \caption{\textbf{Comparison of the Range of Polling Error Estimates from 14 to 28 Days before the Election.} Each point is the largest minus the smallest error estimate for the Linear model (y-axis) and Random Walk Model (x-axis) across the 15 potential cutoff choices between two and four weeks prior to the election. Our proposed model almost always has a narrower range than the Linear Model's range of estimates, indicating that our model is much less sensitive to the subjective analyst decision on which polls to include.}
    \label{fig:error_range_comparison}
\end{figure}

Finally, we compare the variability of the error estimates across different cutoff times $T$. Our RW Model automatically discounts the information from early polls and changes its estimates of error only when fluctuations in preferences are relatively constant ($\gamma_r$ is small). The Linear Model weights all polls equally after removing the linear trend, so it will be more sensitive to the subjective decision of how many polls to include. We compute for each election and model the range of error estimates for cutoffs between 14 and 28 days. Figure~\ref{fig:error_range_comparison} plots the Linear and Random Walk error ranges split by election type, coloured by year, and labeled with the two-letter state abbreviation. We observe that consistently the Linear model estimates are more variable, especially for Senate elections: 75\% of Presidential and 82\% of Senatorial elections have wider ranges of error estimates when using the Linear Model instead of our Random Walk Model.

\section{Conclusion}

Pollsters have tried to adapt to the failures in past elections by changing sampling methodologies and weighting schemes, and certainly future forecasting models will pay more attention to uncertainty and error in polling data \citep{skelley2020change}. How one actually measures that uncertainty and error, especially for use as inputs to forecasts or campaign-related decisions, is an important initial undertaking. Use of simple models is appealing, and assuming constant or linear changes in preferences is in many cases a valid approach. However, over reliance on those models is cautioned against when violations of the model assumptions are not reflected in increased parameter uncertainty and the model conclusions vary based on subjective analyst decisions. The use of the more complex random walk framework allows our model to adapt to those simple assumptions only when they are justified in the data, and, when those assumptions do not hold, our model does not mislabels changes in preferences as polling error. 

Our model better captures turbulent election cycles, has easily interpretable results, and measures polling errors in a way that is robust to subjective inclusion decisions about how many polls to include. When voter preferences are changing or late-breaking news stories alter election dynamics, our model is flexible enough to separate polling error from shifting preferences. Simple models will attribute these changes to directional error or excess variance, reflected in the Linear Models estimates that are 10\% and 25\% more extreme than our model's estimates. This flexibility is further reflected in the much lower variability of estimates when changing how many polls are included. Other methods, with stronger assumptions about how preferences evolve, have marked shifts in estimated bias when more data are included. Moreover, with a single parameter to estimate bias, the results of our model are directly interpretable and do not vary with the actual election result, which eases the use of the model's conclusions in election forecasting and other decision-making contexts where elections results are not yet known. 

As highlighted in Section~\ref{sec:background}, the potential sources of polling errors are varied and difficult to separate. Our method confirms that these additional factors do contribute to excess variability error and, for 2016 and 2020 Presidential elections, notable directional errors that overstated the Democratic candidate's support. However, our model also indicates that the excess variability is much smaller and errors are less extreme than traditional estimates from models that do not separate changes in the target estimand (the electorate's preferences) from variability of the estimator (the poll result). 

One could use our model to try and attribute polling error to poll features, such as sample frame or house effects, by separating $\alpha_r$ into poll-specific indicator variables, or one could attribute polling error to state-level features by replacing $\alpha_r$ with $\vect X_r \beta_r$, where $\vect X_r$ is a vector of election-specific features such as the proportion of white voters without a college degree. 
This decomposition is beyond the scope of this paper as it carries substantial new complications: as the granularity of the decomposition increases, the amount of data available for any particular feature decreases, which in turn requires more complex sharing of information across states and election years (e.g. for contemporaneous elections).

Estimating total polling error and separating bias from variance is a difficult task. With the frequency of systematic errors in recent elections, increased attention has come to this challenge. While these different models often lead to similar conclusions, ensuring that the parameters are easily interpretable and that the conclusions are not sensitive to analyst decisions, such as the inclusion window, is important. The simple interpretation of $\alpha_r$ in our model, as opposed to needing to compute a more complicated logistic transformation, eases the use of our model's results in other scenarios, e.g. election forecasting (which could even be done with our model by simply placing a prior on $v_r$), and enables assessing more complex explanations for polling errors. With a single parameter measuring error independent of the election result, researches can explore how poll or election-level features impact polling errors without ambiguity regarding the outcome of interest.



\section*{Acknowledgement}
We thank participants at the Joint Statistical Meetings and the Junior ISBA workshop for their helpful comments. We also thank D. Sunshine Hillygus and Brian Guay for helpful discussions and reviews of early drafts of this work.  



\printbibliography

@article{Linzer2013election,
    author = { Drew A.   Linzer },
    title = {Dynamic Bayesian Forecasting of Presidential Elections in the States},
    journal = {Journal of the American Statistical Association},
    volume = {108},
    number = {501},
    pages = {124-134},
    year  = {2013},
    publisher = {Taylor & Francis},
    doi = {10.1080/01621459.2012.737735},
    URL = {https://doi.org/10.1080/01621459.2012.737735},
    eprint = {https://doi.org/10.1080/01621459.2012.737735}
}

@article{ShiraniMehr2018biasvariance,
	author = {Houshmand Shirani-Mehr and David Rothschild and Sharad Goel and Andrew Gelman},
	title = {Disentangling Bias and Variance in Election Polls},
	journal = {Journal of the American Statistical Association},
	volume = {113},
	number = {522},
	pages = {607-614},
	year  = {2018},
	publisher = {Taylor & Francis},
	doi = {10.1080/01621459.2018.1448823},
	URL = {https://doi.org/10.1080/01621459.2018.1448823},
	eprint = { https://doi.org/10.1080/01621459.2018.1448823}
}

@article{jennings2018pollerrors,
	Abstract = {Are election polling misses becoming more prevalent? Are they more likely in some contexts than others? Here we undertake an over-time and cross-national assessment of prediction errors in pre-election polls. Our analysis draws on more than 30,000 national polls from 351 general elections in 45 countries between 1942 and 2017. We proceed in the following way. First, building on previous studies, we show how errors in national polls evolve in a structured way over the election timeline. Second, we examine errors in polls in the final week of the election campaign to assess performance across election years. Third, we undertake a pooled analysis of polling errors---controlling for a number of institutional and party features---that enables us to test whether poll errors have increased or decreased over time. We find that, contrary to conventional wisdom, the recent performance of polls has not been outside the ordinary. However, the performance of polls does vary across political contexts and in understandable ways.},
	Author = {Jennings, Will and Wlezien, Christopher},
	Da = {2018/04/01},
	Date-Added = {2020-03-23 19:31:31 +0000},
	Date-Modified = {2020-03-23 19:31:31 +0000},
	Doi = {10.1038/s41562-018-0315-6},
	Id = {Jennings2018},
	%Isbn = {2397-3374},
	Journal = {Nature Human Behaviour},
	Number = {4},
	Pages = {276--283},
	Title = {Election polling errors across time and space},
	Ty = {JOUR},
	Url = {https://doi.org/10.1038/s41562-018-0315-6},
	Volume = {2},
	Year = {2018},
	Bdsk-Url-1 = {https://doi.org/10.1038/s41562-018-0315-6},
	Bdsk-Url-2 = {http://dx.doi.org/10.1038/s41562-018-0315-6}}

@article{gelman2016mythical,
  title={The mythical swing voter},
  author={Gelman, Andrew and Goel, Sharad and Rivers, Douglas and Rothschild, David and others},
  journal={Quarterly Journal of Political Science},
  volume={11},
  number={1},
  pages={103--130},
  year={2016},
  publisher={Now Publishers, Inc.}
}

@article{groves2010total,
  title={Total survey error: Past, present, and future},
  author={Groves, Robert M and Lyberg, Lars},
  journal={Public opinion quarterly},
  volume={74},
  number={5},
  pages={849--879},
  year={2010},
  publisher={Oxford University Press}
}

@article{biemer2010total,
  title={Total survey error: Design, implementation, and evaluation},
  author={Biemer, Paul P},
  journal={Public opinion quarterly},
  volume={74},
  number={5},
  pages={817--848},
  year={2010},
  publisher={Oxford University Press}
}

@book{weisberg2009total,
  title={The total survey error approach},
  author={Weisberg, Herbert F},
  year={2009},
  publisher={University of Chicago Press}
}

@article{mcfarland1981effects,
  title={Effects of question order on survey responses},
  author={McFarland, Sam G},
  journal={Public Opinion Quarterly},
  volume={45},
  number={2},
  pages={208--215},
  year={1981},
  publisher={Oxford University Press}
}

@article{smith1987we,
  title={That which we call welfare by any other name would smell sweeter an analysis of the impact of question wording on response patterns},
  author={Smith, Tom W},
  journal={Public opinion quarterly},
  volume={51},
  number={1},
  pages={75--83},
  year={1987},
  publisher={Oxford University Press}
}

@article{mcdermott2003horserace,
  title={Horserace Polling and Survey Method Effects: an analysis of the 2000 campaign},
  author={McDermott, Monika L and Frankovic, Kathleen A},
  journal={The Public Opinion Quarterly},
  volume={67},
  number={2},
  pages={244--264},
  year={2003},
  publisher={JSTOR}
}

@article{jackman2005pooling,
  title={Pooling the polls over an election campaign},
  author={Jackman, Simon},
  journal={Australian Journal of Political Science},
  volume={40},
  number={4},
  pages={499--517},
  year={2005},
  publisher={Taylor \& Francis}
}

@article{barnes2016reality,
  title={Reality Check: Should We Give Up on Election Polling?},
  author={Barnes, Peter},
  journal={BBC News},
  date={2016-11-16},
  year={2016},
  url={https://www.bbc.com/news/election-us-2016-37949527},
  urldate={2022-01-25}
}

@article{sturgis2016report,
  title={Report of the inquiry into the 2015 British general election opinion polls},
  author={Sturgis, Patrick and Baker, Nick and Callegaro, Mario and Fisher, Stephen and Green, Jane and Jennings, Will and Kuha, Jouni and Lauderdale, Ben and Smith, Patten},
  year={2016},
  journal={NCRM, British Polling Council, Market Research Society},
  url={https://eprints.soton.ac.uk/390588/1/Report_final_revised.pdf}
}

@article{aapor2017evaluation,
  title={An evaluation of 2016 election polls in the US},
  author={Kennedy, Courtney and Blumenthal, Mark and Clement, Scott and Clinton, Joshua D and Durand, Claire and Franklin, Charles and McGeeney, Kyley and Miringoff, Lee and Olson, Kristen and Rivers, Douglas and Saad, Lydia and Witt, G. Evans and Wlezien, Christopher},
  year={2017},
  url={https://www.aapor.org/Education-Resources/Reports/An-Evaluation-of-2016-Election-Polls-in-the-U-S.aspx},
  urldate={2022-01-25},
    journal={American Association for Public Opinion Research},
}

@article{aapor2021evaluation,
  title={Task Force on 2020 Pre-Election Polling: An Evaluation of the 2020 General Election Polls},
  author={Clinton, Josh and Agiesta, Jennifer and Brenan, Megan and Burge, Camille and Connelly, Marjorie and Edwards-Levy, Ariel and Fraga, Bernard and Guskin, Emily and Hillygus, D Sunshine and Jackson, Chris and Jones, Jeff and Keeter, Scott and Khanna, Kabir and Lapinski, John and Saad, Lydia and Shaw, Daron and Smith, Andrew and Wilson, David and Wlezien, Christopher},
  journal={American Association for Public Opinion Research},
  year={2021}
}

@article{kennedy2018evaluation,
  title={An evaluation of the 2016 election polls in the United States},
  author={Kennedy, Courtney and Blumenthal, Mark and Clement, Scott and Clinton, Joshua D and Durand, Claire and Franklin, Charles and McGeeney, Kyley and Miringoff, Lee and Olson, Kristen and Rivers, Douglas and Saad, Lydia and Witt, G. Evans and Wlezien, Christopher},
  journal={Public Opinion Quarterly},
  volume={82},
  number={1},
  pages={1--33},
  year={2018},
  publisher={Oxford University Press US}
}

@incollection{jackson2018rise,
  title={The rise of poll aggregation and election forecasting},
  author={Jackson, Natalie},
  booktitle={The Oxford Handbook of polling and survey methods},
  year={2018},
  publisher = {Oxford University Press},
  editor={Atkeson, L.R. and Alvarez, R.M.},
  address = {Oxford}
}

@article{heidemanns2020updated,
  title={An updated dynamic Bayesian forecasting model for the US presidential election},
  author={Heidemanns, Merlin and Gelman, Andrew and Morris, G Elliott},
  journal={Harvard Data Science Review},
  volume={2},
  number={4},
  year={2020},
  publisher={PubPub}
}

@article{stoetzer2019forecasting,
  title={Forecasting elections in multiparty systems: a Bayesian approach combining polls and fundamentals},
  author={Stoetzer, Lukas F and Neunhoeffer, Marcel and Gschwend, Thomas and Munzert, Simon and Sternberg, Sebastian},
  journal={Political Analysis},
  volume={27},
  number={2},
  pages={255--262},
  year={2019},
  publisher={Cambridge University Press}
}

@article{chen2022polls,
  title={Polls, Context, and Time: A Dynamic Hierarchical Bayesian Forecasting Model for US Senate Elections},
  author={Chen, Yehu and Garnett, Roman and Montgomery, Jacob M},
  year={2022},
  journal={Political Analysis},
  pages={1--21},
  publisher={Cambridge University Press}
}

@article{abramowitz2008forecasting,
  title={Forecasting the 2008 presidential election with the time-for-change model},
  author={Abramowitz, Alan I},
  journal={PS: Political Science \& Politics},
  volume={41},
  number={4},
  pages={691--695},
  year={2008},
  publisher={Cambridge University Press}
}

@article{erikson2008leading,
  title={Leading economic indicators, the polls, and the presidential vote},
  author={Erikson, Robert S and Wlezien, Christopher},
  journal={PS: Political Science \& Politics},
  volume={41},
  number={4},
  pages={703--707},
  year={2008},
  publisher={Cambridge University Press}
}

@Misc{rstan,
    title = {{RStan}: the {R} interface to {Stan}},
    author = {{Stan Development Team}},
    note = {R package version 2.21.2},
    year = {2020},
    url = {http://mc-stan.org/},
  }

@article{caughey_warshaw_2015, title={Dynamic Estimation of Latent Opinion Using a Hierarchical Group-Level IRT Model}, volume={23}, DOI={10.1093/pan/mpu021}, number={2}, journal={Political Analysis}, publisher={Cambridge University Press}, author={Caughey, Devin and Warshaw, Christopher}, year={2015}, pages={197–211}}

@article{walther2015picking,
  title={Picking the winner (s): Forecasting elections in multiparty systems},
  author={Walther, Daniel},
  journal={Electoral Studies},
  volume={40},
  pages={1--13},
  year={2015},
  publisher={Elsevier}
}

@article{jackson2020uncertainty,
  title={Poll-Based Election Forecasts Will Always Struggle With Uncertainty},
  author={Jackson, Natalie},
  journal={Center for Politics},
  date={2020-08-06},
  year={2020},
  url={https://centerforpolitics.org/crystalball/articles/poll-based-election-forecasts-will-always-struggle-with-uncertainty/},
  urldate={2022-01-31}
}

@article{huang2009beyond,
  title={Beyond the battlegrounds? Electoral college strategies in the 2008 presidential election},
  author={Huang, Taofang and Shaw, Daron},
  journal={Journal of Political Marketing},
  volume={8},
  number={4},
  pages={272--291},
  year={2009},
  publisher={Taylor \& Francis}
}

@article{skelley2020change, title={What Pollsters Have Changed Since 2016 — And What Still Worries Them About 2020.}, 
    url={https://fivethirtyeight.com/features/what-pollsters-have-changed-since-2016-and-what-still-worries-them-about-2020/},
    journal={FiveThirtyEight},
    publisher={FiveThirtyEight}, 
    author={Skelley, Geoffrey and Rakich, Nathaniel},
    year={2020},
    month={10}}

@article{sunshine2011evolution,
    author = {Hillygus, D. Sunshine},
    title = "{The Evolution of Election Polling in the United States}",
    journal = {Public Opinion Quarterly},
    volume = {75},
    number = {5},
    pages = {962-981},
    year = {2011},
    month = {12},
    issn = {0033-362X},
    doi = {10.1093/poq/nfr054},
    url = {https://doi.org/10.1093/poq/nfr054},
    eprint = {https://academic.oup.com/poq/article-pdf/75/5/962/5180996/nfr054.pdf},
}

@article{pickup2008campaign,
  title={Campaign trial heats as election forecasts: Measurement error and bias in 2004 presidential campaign polls},
  author={Pickup, Mark and Johnston, Richard},
  journal={International Journal of Forecasting},
  volume={24},
  number={2},
  pages={272--284},
  year={2008},
  publisher={Elsevier}
}

@article{pickup2007campaign,
  title={Campaign trial heats as electoral information: evidence from the 2004 and 2006 Canadian federal elections},
  author={Pickup, Mark and Johnston, Richard},
  journal={Electoral Studies},
  volume={26},
  number={2},
  pages={460--476},
  year={2007},
  publisher={Elsevier}
}

@article{campbell1990trial,
  title={Trial-heat forecasts of the presidential vote},
  author={Campbell, James E and Wink, Kenneth A},
  journal={American Politics Quarterly},
  volume={18},
  number={3},
  pages={251--269},
  year={1990},
  publisher={Sage Publications Sage CA: Thousand Oaks, CA}
}

@book{prado2010time,
  title={Time series: modeling, computation, and inference},
  author={Prado, Raquel and West, Mike},
  year={2010},
  publisher={Chapman and Hall/CRC}
}

@article{fey1997stability,
  title={Stability and coordination in Duverger's law: A formal model of preelection polls and strategic voting},
  author={Fey, Mark},
  journal={American Political science review},
  volume={91},
  number={1},
  pages={135--147},
  year={1997},
  publisher={Cambridge University Press}
}

@article{levine2007paradox,
  title={The paradox of voter participation? A laboratory study},
  author={Levine, David K and Palfrey, Thomas R},
  journal={American political science Review},
  volume={101},
  number={1},
  pages={143--158},
  year={2007},
  publisher={Cambridge University Press}
}

@inproceedings{hillygus2016polling,
  title={Polling in the United States},
  author={Hillygus, D SUNSHINE and Guay, Brian},
  booktitle={The Seminar Magazine issue on ‘Measuring Democracy},
  volume={684},
  year={2016}
}

@article{rothschild2009forecasting,
  title={Forecasting elections: Comparing prediction markets, polls, and their biases},
  author={Rothschild, David},
  journal={Public Opinion Quarterly},
  volume={73},
  number={5},
  pages={895--916},
  year={2009},
  publisher={Oxford University Press}
}

@article{clinton2013robo,
  title={Robo-Polls: Taking Cues from Traditional Sources?},
  author={Clinton, Joshua D and Rogers, Steven},
  journal={PS: Political Science \& Politics},
  volume={46},
  number={2},
  pages={333--337},
  year={2013},
  publisher={Cambridge University Press}
}

@article{sturgis2018assessment,
  title={An assessment of the causes of the errors in the 2015 UK general election opinion polls},
  author={Sturgis, Patrick and Kuha, Jouni and Baker, Nick and Callegaro, Mario and Fisher, Stephen and Green, Jane and Jennings, Will and Lauderdale, Benjamin E and Smith, Patten},
  journal={Journal of the Royal Statistical Society: Series A (Statistics in Society)},
  volume={181},
  number={3},
  pages={757--781},
  year={2018},
  publisher={Wiley Online Library}
}

@article{bon2019polling,
  title={Polling bias and undecided voter allocations: US presidential elections, 2004--2016},
  author={Bon, Joshua J and Ballard, Timothy and Baffour, Bernard},
  journal={Journal of the Royal Statistical Society: Series A (Statistics in Society)},
  volume={182},
  number={2},
  pages={467--493},
  year={2019},
  publisher={Wiley Online Library}
}

@article{pasek2015predicting,
  title={Predicting elections: Considering tools to pool the polls},
  author={Pasek, Josh},
  journal={Public Opinion Quarterly},
  volume={79},
  number={2},
  pages={594--619},
  year={2015},
  publisher={Oxford University Press US}
}

@article{erikson1999presidential,
  title={Presidential polls as a time series: the case of 1996},
  author={Erikson, Robert S and Wlezien, Christopher},
  journal={Public opinion quarterly},
  pages={163--177},
  year={1999},
  publisher={JSTOR}
}

@article{wlezien2002timeline,
  title={The timeline of presidential election campaigns},
  author={Wlezien, Christopher and Erikson, Robert S},
  journal={The Journal of Politics},
  volume={64},
  number={4},
  pages={969--993},
  year={2002},
  publisher={Blackwell Publishing, Inc.}
}

@article{green1999tracking,
  title={Tracking opinion over time: A method for reducing sampling error},
  author={Green, Donald P and Gerber, Alan S and De Boef, Suzanna L},
  journal={Public Opinion Quarterly},
  pages={178--192},
  year={1999},
  publisher={JSTOR}
}

\end{document}